\documentclass{webofc}
\usepackage[varg]{txfonts}   
\usepackage{braket}
\usepackage{multirow}
\usepackage{dsfont}
\usepackage{mathtools}
\usepackage{xcolor}
\usepackage[normalem]{ulem}

\begin{document}
\title{$SU(2N_F)$ symmetry of confinement in QCD and its observation
at high temperature.}
%
%

\author{\firstname{L. Ya. } \lastname{Glozman}\inst{1}\fnsep\thanks{\email{leonid.glozman@uni-graz.at}} 
}

\institute{Institute of Physics, University of Graz, A-8010 Graz, Austria
          }

\abstract{ In this talk we first overview  lattice results that have led to the  observation of  new $SU(2)_{CS}$ and $SU(2N_F)$ symmetries upon artificial
truncation of the near-zero modes of the Dirac operator at zero temperature and at high temperature without any truncation. These symmetries are larger than the chiral symmetry of the QCD Lagrangian and contain chiral symmetries $SU(N_F)_L \times SU(N_F)_R$ and $U(1)_A$ as subgroups. In addition to the standard chiral transformations the $SU(2)_{CS}$ and $SU(2N_F)$ transformations mix the right- and left-handed components of the quark fields. It is a symmetry of the confining
chromo-electric interaction while the chromo-magnetic interaction manifestly
breaks it. Emergence of these symmetries  upon truncation of the
near-zero modes of the Dirac operator at T=0 means that all effects of the chromo-magnetic interaction are located exclusively in the near-zero modes,
while confining chromo-electric interaction is distributed among all modes.
Appearance of these symmetries at high T, where the temperature suppresses the
near-zero modes, has  radical implications because these symmetries are
incompatible with the asymptotically free deconfined quarks at increasing temperature. The elementary objects in the high-temperature phase of QCD
should be quarks bound by the pure chromo-electric field that is not
accompanied by the chromo-magnetic effects.
}
\maketitle
\section{Introduction}
\label{intro}
The QCD Lagrangian   with $N_F$ massles quarks

\begin{equation}
{\cal {L}} = \bar{\Psi}(x)( i \gamma_\mu D_\mu ) \Psi(x) -\frac{1}{2}
Tr(G^{\mu\nu}G_{\mu\nu}),
\label{lag}
\end{equation}

\noindent
has the chiral symmetry: 

\begin{equation}
U(N_F)_L \times U(N_F)_R =
SU(N_F)_L \times SU(N_F)_R \times U(1)_A \times U(1)_V.
\label{qcdsymm}
\end{equation}

\noindent
The $U(1)_V$ symmetry is responsible for the vector current conservation and
is irrelevant
to our subject. The $U(1)_A$ symmetry is an invariance  upon the axial flavor-neutral transformations

\begin{equation}
\Psi(x) \rightarrow e^{i\alpha \gamma_5}\Psi(x); ~~~~
\bar \Psi(x) \rightarrow \bar \Psi(x) e^{i\alpha \gamma_5}.
\label{u1a}
\end{equation}

\noindent
The $SU(N_F)_L \times SU(N_F)_R$ chiral symmetry is an invariance
under independent flavor  $SU(N_F)$ rotations of the left- and right-handed
components of quarks. These transformations contain the flavor (isospin for $N_F=2$) rotations  as well as  the axial flavor transformations

\begin{equation}
\Psi(x) \rightarrow e^{i \gamma_5 \frac{\vec{\lambda}\cdot \vec{\alpha}}
{2}}
\Psi(x); ~~~~
\bar \Psi(x) \rightarrow \bar \Psi(x) e^{i\gamma_5 \frac{\vec{\lambda}\cdot \vec{\alpha}}
{2}},
\label{u2a}
\end{equation}

\noindent
where $\vec{\lambda}$ are $SU(N_F)$ generators.

The  $U(1)_A$ symmetry is broken anomalously, which is due to
a noninvariance of the integration measure in the functional integral under a local
 $U(1)_A$ transformation \cite{FU}.
The $ SU(N_F)_A $ "symmetry" (the transformations  (\ref{u2a}) do not form a closed
subgroup of the chiral group)  is broken spontaneously (dynamically).
The ground state, the vacuum, is not invariant under
the transformation (\ref{u2a}). This  is encoded in
the  quark condensate, $<0| \bar\Psi(x) \Psi(x)|0 > \neq 0$.

The quark condensate of the vacuum is connected to the density of the
near-zero modes  of the Euclidean Dirac operator
via the Banks-Casher relation \cite{BC}

\begin{equation}
\lim_{m \rightarrow 0} <0|\bar \Psi(x) \Psi(x)|0> = -\pi \rho(0)~.
\label{bc}
\end{equation}

Historically it was believed that the mass  of hadrons
in the $u,d$ quark sector is generated via the spontaneous breaking
of chiral symmetry. For example, within the linear sigma-model the
nucleon acquires its mass via its coupling with the chiral order
parameter. Similar, within the Nambu and Jona-Lasinio model it is
the vacuum fermion  condensate that is responsible for a generation of a large 
mass of initially massless fermions. This view was one of the reasons
to assume that at high temperatures, where chiral symmetry is restored,
there should appear the quark-gluon plasma phase, contrary to the hadron
phase at low temperatures.

One can artificially restore the flavor-nonsinglet chiral symmetry
by removing  the quark condensate by hands, i.e. via subtraction  of the near-zero modes of the Dirac operator on the lattice \cite{LS}.
It turned out that  the nucleon, the rho-meson and some other hadrons, 
except for the pion, survive this "unbreaking" of the chiral symmetry and their mass remains large \cite{GLS}. This tells that while the
chiral symmetry breaking in the vacuum is important for the eventual shape of the hadron spectra, the chiral symmetry breaking is not
the main mechanism of the mass generation of hadrons such as the $\rho$-meson
or the nucleon. Still it was unclear what happens with the $U(1)_A$ symmetry
upon truncation of the near-zero modes. This was a motivation for a systematic study of the hadron spectra 
 upon truncation of the near-zero modes within $N_F=2$
dynamical calculations with the chirally-invariant overlap Dirac operator\cite{D1,D2,D3,D4} . This study has
led to discovery of new $SU(2)_{CS}$ and $SU(4)$ symmetries that are larger
than the chiral symmetry of the QCD Lagrangian and that allow mixing of the
left- and right-handed components of the quark field \cite{G1,G2}.

Below we consider observation of these symmetries and their implication.
Then we report  recent results on  these symmetries at high temperature without any truncation \cite{R} and their consequences for the nature of
the strongly interacting matter at high T.

\section{Truncation of the near-zero modes and naive expectations}

 The hermitian Euclidean Dirac operator, $i \gamma_\mu D_\mu$, where
 $D_\mu = \partial_\mu + i g\frac{t^a}{2} A^a_\mu$
 has in a finite volume $V$ a discrete spectrum with real eigenvalues $\lambda_n$:

\begin{equation}
i \gamma_\mu D_\mu  \Psi_n(x) = \lambda_n \Psi_n(x).
\label{ev}
\end{equation}

\noindent
We subtract from the  quark propagator $S_{Full}$ the $k$ lowest  eigenmodes
of the Dirac operator

\begin{equation}
S(x,y) =S_{Full}(x,y)-
  \sum_{n=1}^{k}\,\frac{1}{\lambda_n + im}\, \Psi_n (x) \Psi_n^\dagger(y).
\label{prop}
\end{equation}

\noindent
 Then we apply standard procedures to extract
hadron spectra using the variational approach.  We perform $N_F =2$ dynamical lattice calculations with the overlap Dirac operator with the gauge
configurations generated by the JLQCD collaboration \cite{JLQCD}, for  details
see refs. \cite{D1,D2}.

Upon truncation of the near-zero Dirac modes, that are 
connected to the quark condensate of the vacuum,
we can expect chiral $SU(2)_L \times SU(2)_R$ symmetry
in correlators. If the hadron states survive this truncation, i.e. an
exponential decay of the correlators is observed,
then we can expect a mass degeneracy of chiral partners. The chiral partners
of  the $J=1$ mesons on Fig. 1 are linked by the red arrows.

\begin{figure}
\centering
\includegraphics[angle=0,width=0.7\linewidth]{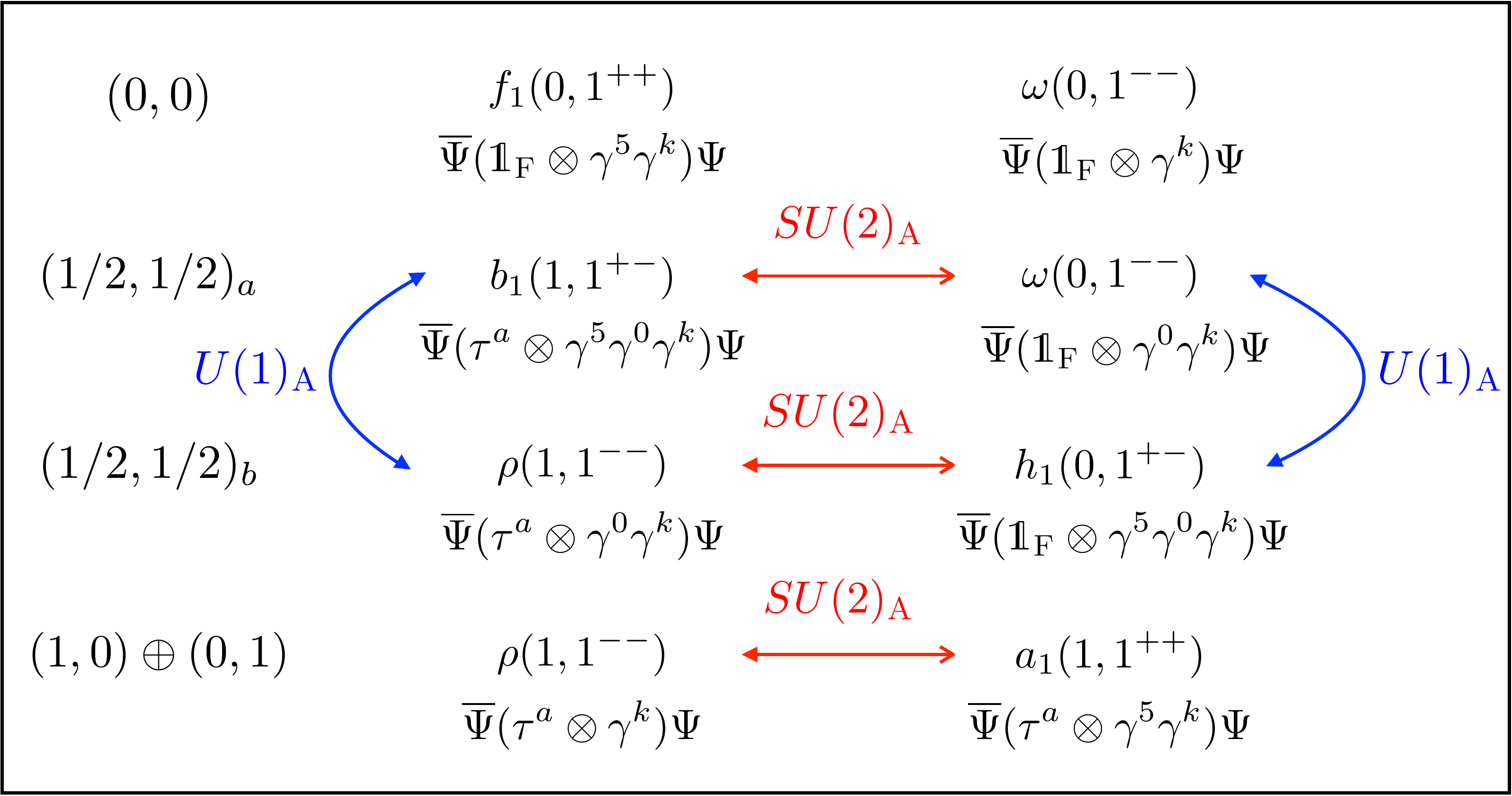}
\caption{$SU(2)_L \times SU(2)_R$  classification of the $J=1$
meson operators. Operators that are connected by the 
$SU(2)_L \times SU(2)_R$ and $U(1)_A$ transformations are connected
by the red and blue arrows, respectively.}
\label{fig-1}
\end{figure}

The $U(1)_A$ transformations connect other operators that are linked by the
blue arrows. If the whole chiral symmetry of QCD
$SU(2)_L \times SU(2)_R \times U(1)_A$ is restored, then we should expect
a degeneracy of four mesons from the $(1/2,1/2)_a$ and $(1/2,1/2)_b$
chiral representations, and in addition a degeneracy
of the $\rho$ and $a_1$ mesons from the $(1,0)+(0,1)$ chiral representation.
\footnote{In the chirally symmetric world there are two independent and orthogonal
$\rho$-mesons that belong to two different chiral representations. In a
world with the chiral symmetry breaking two different $\rho$ operators couple to one and the same $\rho$-meson, because its wave function is a mixture
of two chiral representations.}

 Consequently, given
only the $SU(2)_L \times SU(2)_R \times U(1)_A$ chiral symmetry we
should expect a degeneracy of all mesons connected by the arrows on Fig. 1.

\section{Results}

Upon truncation of some amount of the low-lying Dirac modes
a very clean exponential decay of all $J=1$ correlators is observed,
which means that there are physical states. For the pion it is the other way round.
After truncation of  a few modes  there is no exponential decay of the pion correlator, which implies that the pion does not survive truncation. The quark
condensate is crucially important for the existence of the pseudo Goldstone
bosons.
After truncation the exponential decay of all $J=1$ correlators
is much cleaner than in the untruncated (real) world. The reason
for this is intuitively clear: After truncation there are no pion
fluctuations in the system. We can conclude that mesons (which are
bound states now) survive the truncation.

\begin{figure}
\centering 
\includegraphics[width=0.55\linewidth]{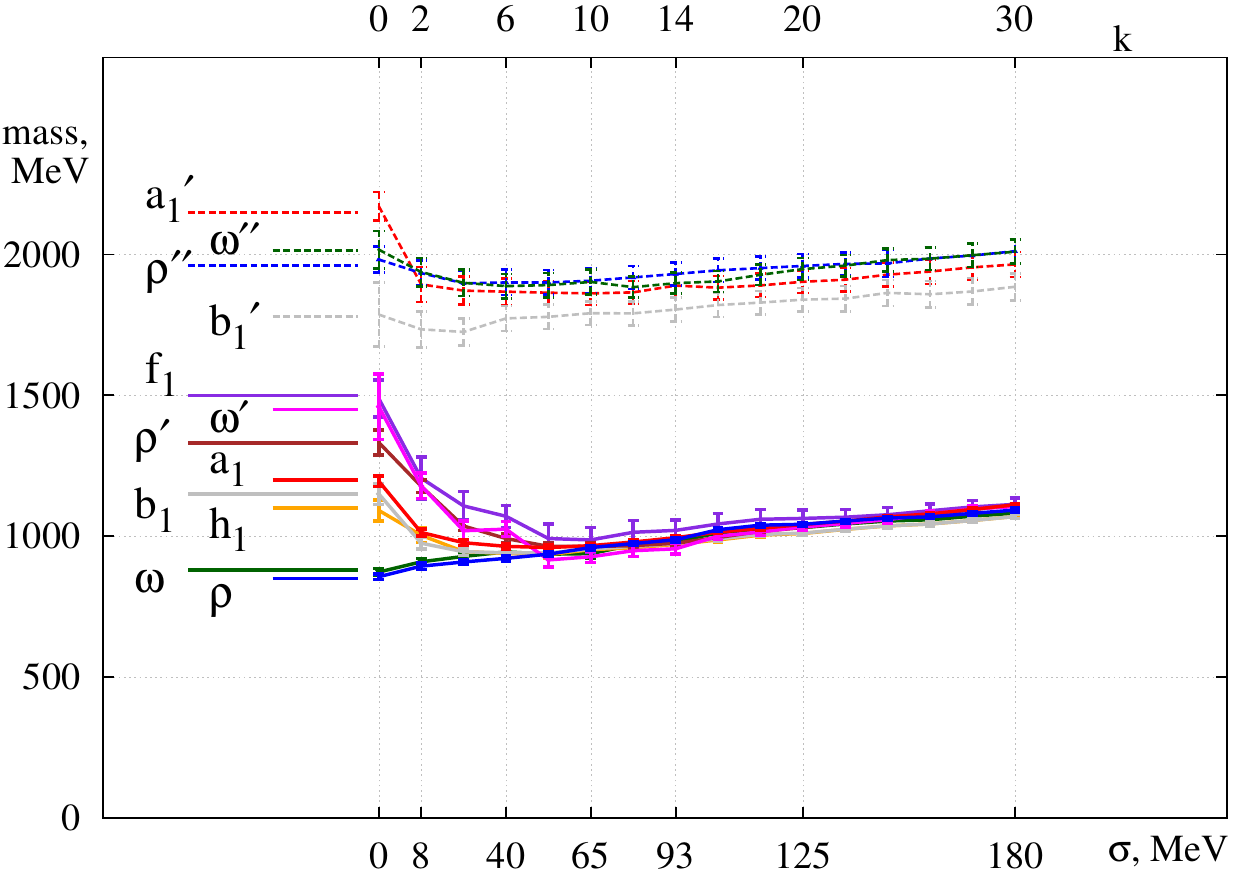}
\label{mass}
\caption{$J=1$ meson mass evolution as a function of the truncation
number $k$. $\sigma$ shows energy gap in the Dirac spectrum.} 
\end{figure}
\noindent

The evolution of meson masses upon truncation of $k$ lowest eigenmodes
is shown in Fig. 2. Chiral (and $U(1)_A$) symmetry restoration happens
at k=10-20. At the same time the hadron mass is large, of the order
of 1 GeV. In this regime the mass generation is obviously not
connected to the chiral symmetry breaking and to the quark condensate.
We conclude that while the spontaneous chiral symmetry  breaking
is important for hadron mass in the real world and for the shape
of observed hadron spectra, the hadron mass arises mostly not from
the chiral symmetry breaking.

Starting with k=10-20 we clearly see a larger degeneracy than the chiral 
$SU(2)_L \times SU(2)_R \times U(1)_A$ symmetry of the QCD
Lagrangian. 
 This implies that there is a symmetry in the system
that is higher than  $SU(2)_L \times SU(2)_R \times U(1)_A$.
What does it mean?! The same results persist for the
$J=2$ mesons \cite{D3} and baryons \cite{D4}.

\section{ $SU(2)_{CS}$ and $SU(4)$ symmetries}

 First we need to understand  what
symmetry group   corresponds to the observed degeneracy \cite{G1}.

Given the standard  quantum numbers we
can construct basis vectors for all irreducible
representations of the chiral group  in Fig. 1. 

\small
{\bf (i)~~~ (0,0):}

\begin{equation}
|(0,0); \pm; J \rangle = \frac{1}{\sqrt 2} |\bar R R \pm \bar L L\rangle_J.
\end{equation}

{\bf (ii)~~~ $(1/2,1/2)_a$ and $(1/2,1/2)_b$:} 

\begin{equation}
|(1/2,1/2)_a; +;I=0; J \rangle = \frac{1}{\sqrt 2} |\bar R L + \bar L R\rangle_J,
\end{equation}

\begin{equation}
|(1/2,1/2)_a; -;I=1; J \rangle = \frac{1}{\sqrt 2} |\bar R \vec \tau L -
\bar L \vec \tau R\rangle_J,
\end{equation} 

\medskip

\begin{equation}
|(1/2,1/2)_b; -;I=0; J \rangle = \frac{1}{\sqrt 2} |\bar R L - \bar L R\rangle_J,
\end{equation}

\begin{equation}
|(1/2,1/2)_b; +;I=1; J \rangle = \frac{1}{\sqrt 2} |\bar R \vec \tau L +
\bar L \vec \tau R\rangle_J.
\end{equation}

{\bf (iii)~~~ (0,1)$\oplus$(1,0):} 

\begin{equation}
|(0,1)+(1,0); \pm; J \rangle = \frac{1}{\sqrt 2} |\bar R \vec \tau R 
\pm \bar L  \vec \tau L \rangle_J,
\end{equation}

 The new symmetry transformations must
connect all these basis vectors. The latter 
can be achieved if one allows a mixing of
 the left- and right-handed quarks. Consequently, the
 symmetry group  that we are looking for must contain as a subgroup the
$SU(2)_{CS}$ {\it chiralspin} rotations that act on the
following fundamental doublets:  

\begin{equation}
\textsc{U} =\begin{pmatrix} u_L \\ u_R\end{pmatrix}, ~~~~~~~
\textsc{D} =\begin{pmatrix} d_L \\ d_R \end{pmatrix}. 
\end{equation}
An imaginary  three-dimensional  space where these rotations are
performed is refered to as the {\it chiralspin}  space. The   rotations
in the chiralspin space mix the right- and left-handed components of the fermion fields independently from the quark flavor.
It is similar to the well familiar isospin space: Rotations in the isospin space mix particles with different
electric charges.

We can construct  explicit representations of the $SU(2)_{CS}$ 
transformations that act on Dirac bispinors \cite{G2}.
Then the $SU(2)_{{CS}}$ chiralspin rotations are generated through 
\begin{equation}
\boldsymbol{\Sigma} = \{ \gamma^k, -i \gamma^5 \gamma^k, \gamma^5 \} \;,
\end{equation}
\noindent
where $\gamma^k, ~k=1,...,4$ is any of the Dirac matrices in Euclidean
space, 
\begin{equation}
 \gamma^i \gamma^j + \gamma^j \gamma^i = 2\delta_{ij}; ~~~ 
\gamma^5 = \gamma^1 \gamma^2 \gamma^3 \gamma^4. 
\end{equation}
\noindent
The $SU(2)$ algebra
 \begin{equation}
[\Sigma^\alpha,\Sigma^\beta] = 2 i \epsilon^{\alpha \beta \gamma} \, \Sigma^\gamma \; ,
\end{equation}
\noindent
 is satisfied with any  $k=1,...,4$ in (15). Different $k$ define
 different $SU(2)_{CS}$ irreducible representations of dim=2.
The Dirac spinor transforms
under a given  $SU(2)_{CS}$ representation as
\begin{equation}
\label{V-def}
  \Psi \rightarrow  \Psi^\prime = e^{i  {\bf{\varepsilon} \cdot \bf{\Sigma}}/{2}} \Psi  \; .
\end{equation}
Upon the $SU(2)_{CS}$ rotations of the Dirac spinors with $k=4$ the operators from
the Fig. 1 transform as triplets or singlets, see Fig. 3.
The $SU(2)_{CS}$ symmetry requires that mesons within triplets should be
degenerate.

If one combines the $SU(2)_{CS}$ and the flavor $SU(2)_F$ group into one
larger group then one arrives at the $SU(4)$ group with the fundamental
vector

\begin{equation}
\Psi =\begin{pmatrix} u_{\textsc{L}} \\ u_{\textsc{R}}  \\ d_{\textsc{L}}  \\ d_{\textsc{R}} \end{pmatrix}. 
\end{equation}
The $SU(4)$ group  contains at the same time $SU(2)_L \times SU(2)_R$
and $SU(2)_{CS} \supset U(1)_A$ as subgroups  
and has the following set of generators:
$$ \{(\tau^a \otimes {1}_D), (1_F \otimes \Sigma^i), (\tau^a \otimes \Sigma^i) \} \; .$$
The global  $SU(4)$ transformations of the Dirac spinor
are defined through
\begin{equation}
\label{W-def}
\Psi \rightarrow  \Psi^\prime = e^{i \bf{\epsilon} \cdot \bf{T}/2} \Psi\; .
\end{equation}
The $SU(4)$ transformations connect all operators from the 15-plet,
see Fig. 3.

For related studies see also \cite{TDC,S}.

\begin{figure}
\centering
\includegraphics[angle=0,width=0.85\linewidth]{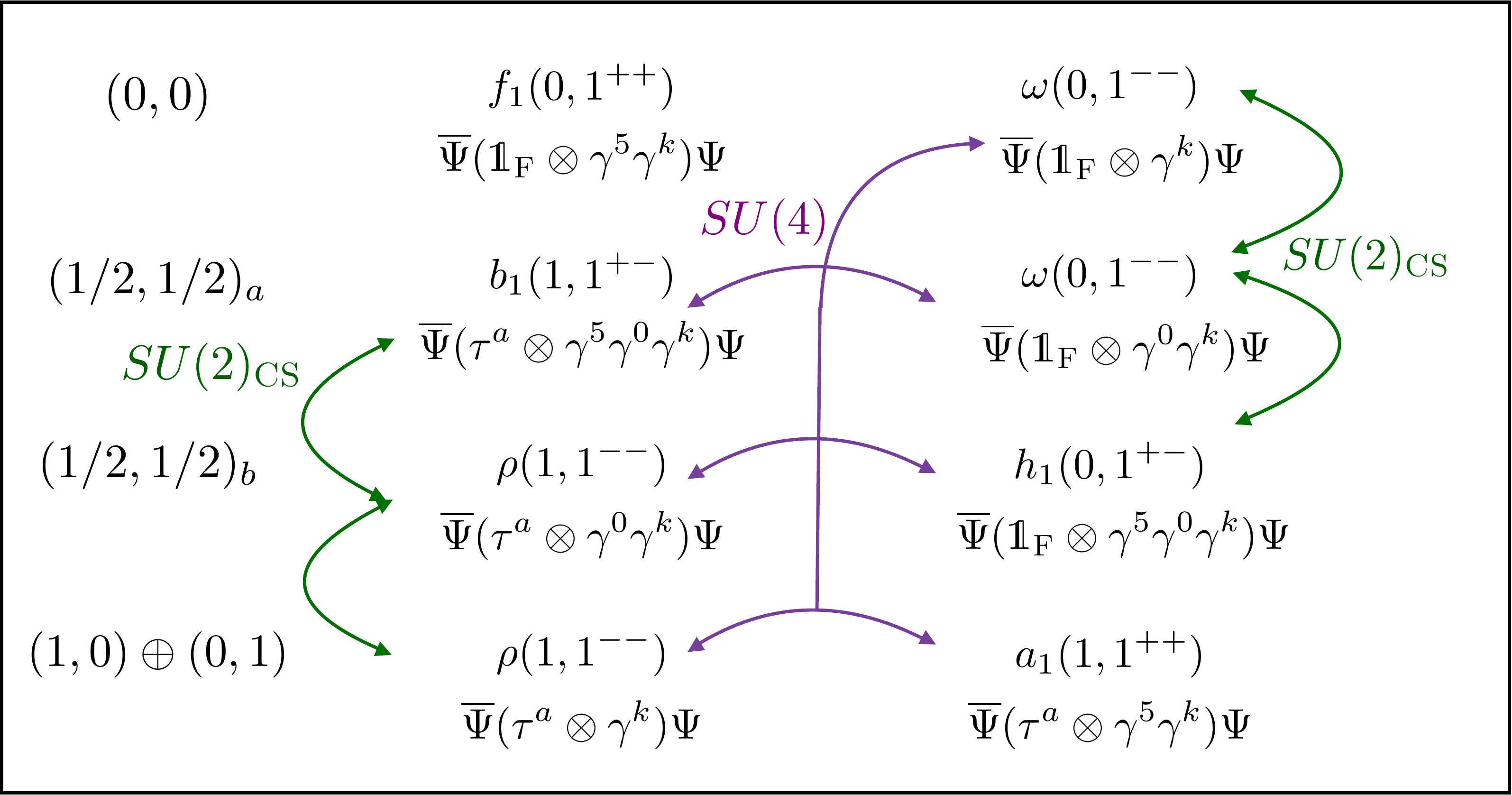}
\caption{ The green arrows connect operators that belong to the
$SU(2)_{CS}$ triplets. The $f_1$ and $a_1$ operators are the 
$SU(2)_{CS}$ singlets. The purple arrows show the $SU(4)$ 15-plet.
The $f_1$ operator is a singlet of $SU(4)$.}
\end{figure}

\section{Chromo-electric versus chromo-magnetic interactions in QCD}

While the QCD Lagrangian does not have the $SU(2)_{CS}$ and $SU(4)$
symmetries one clearly observes emergence of these symmetries upon
elimination of the near-zero modes. Their emergence tells that there
is    some  dynamics in QCD that is $SU(2)_{CS}-$ and $SU(4)-$symmetric.
At the same time there is another dynamics that breaks both symmetries and the
latter dynamics is intrinsically connected only to the low-lying modes
of the Dirac operator.

Consider the  interaction part of the QCD Lagrangian:

\begin{equation}
\label{Lagrangian}
 \overline{\Psi}  i \gamma^{\mu} D_{\mu} \Psi = \overline{\Psi}  i \gamma^0 D_0  \Psi 
  + \overline{\Psi}  i \gamma^i D_i  \Psi\; . 
\end{equation}

\noindent
The first (temporal) term is an interaction of the quark charge density 
$\rho(x) = \bar \Psi (x) \gamma^0 \Psi(x) = \Psi (x)^\dagger \Psi(x)$ with the chromo-electric  
part of the gluonic field. The second (spatial) term contains a quark kinetic term
and  an
interaction of the spatial current density with the chromo-magnetic field.
The temporal part is invariant under any unitary transformation that can
be defined in the Dirac spinor space. In particular it is invariant under the
$SU(2)_{CS}$ and $SU(4)$ transformations.
At the same time 
the kinetic quark term and the magnetic part of the interaction Lagrangian do not admit these higher
symmetries (because the magnetic interaction explicitly distinguishes  the left and the right and is not invariant upon the $L \leftrightarrow R$ operation) and are invariant only with respect to global $SU(N_F)_L \times SU(N_F)_R \times U(1)_A$ chiral transformations \cite{G2}.
Consequently QCD has,
given the $U(1)_A$ anomaly, only the $SU(N_F)_L \times SU(N_F)_R$ chiral symmetry.

This symmetry classification of different parts of the interaction 
Lagrangian together with emergence of the $SU(2)_{CS}$ and $SU(4)$
symmetries upon truncation of the near-zero modes implies that effect
of the chromo-magnetic interaction in QCD is located exclusively in
the near-zero modes, while confining chromo-electric interaction is
distributed among all modes of the Dirac operator. Obviously some not
yet known microscopic
dynamics should responsible for this phenomenon.

To summarize this section, the chromo-electric interaction, that is  responsible for confinement in QCD, is $SU(2)_{CS}$ and $SU(2N_F)$
symmetric. One can speculate that the confining dynamics is
realized via a dynamical chromo-electric string \cite{G1,G2}.

\section{Zero modes and $SU(2)_{CS}$}
The  QCD Lagrangian is not invariant under 
$SU(2)_{CS}$ and $SU(4)$ transformations. 
This is
because the Dirac operator does not commute with all $SU(2)_{CS}$
generators.  
The $SU(2)_{CS}$ and $SU(4)$ symmetries are obtained
in lattice simulations upon subtraction of the near-zero modes of the Dirac
operator. This hints that dynamically the $SU(2)_{CS}$ breaking in
QCD might be related to topology and zero modes.

Consider the zero modes of the Dirac equation, 

\begin{equation}
 \gamma_\mu D_\mu  \Psi_0(x) = 0.
\label{dir}
\end{equation}

\noindent
With  the standard antiperiodic boundary conditions for the quark
field along the time direction, the
zero modes are solutions of the Dirac equation with a gauge
configuration of a nonzero global topological charge.  
The difference of numbers of
the left-handed
and right-handed zero modes  is fixed, according to the Atiyah-Singer
theorem, by the global topological charge $Q$ of the gauge configuration:

\begin{equation}
 n_L - n_R = Q.
\label{AZ}
\end{equation}

\noindent
Some $SU(2)_{CS}$ transformations rotate
the right-handed spinor into the left-handed one and vice versa.
The $SU(2)_{CS}$ symmetry is possible only if the amount of the
left-handed and right-handed zero modes is the same.
Consequently,  within a fixed $Q \neq 0$ sector the $SU(2)_{CS}$
should be broken. 

While the exact zero modes of the Dirac operator are irrelevant
for observables in the thermodynamical limit, it might happen
that via  fluctuations of the gluonic field this asymmetry between
the left and the right related to topology becomes a property of the near-zero modes.

\section{Observation of approximate $SU(2)_{CS}$ and $SU(4)$ symmetries
at high temperatures \cite{R}}

So far we have discussed $SU(2)_{CS}$ and $SU(4)$ symmetries
that emerge upon artificial truncation of the near-zero Dirac modes in $T=0$
calculations. The near-zero modes of the Dirac operator are naturally
suppressed at high temperature. It is well established on the lattice
that above the 
critical temperature the chiral $SU(2)_L \times SU(2)_R$ symmetry gets restored
and the quark condensate of the vacuum vanishes, i.e. a density of the quasi-zero modes is zero. Even more, there are clear indications from the
 simulations with the chirally invariant Dirac operator that above
$T_c$ also  $U(1)_A$ is restored and a gap opens in the Dirac spectrum
\cite{JLQCD1,JLQCD2}. Then, one can expect that at high temperatures
not only the $SU(2)_L \times SU(2)_R \times U(1)_A$ symmetry of the
QCD Lagrangian is manifest but also the $SU(2)_{CS}$ and $SU(4)$ symmetries
emerge with  far reaching consequences \cite{G3}.

Given this expectation spatial ($z$-direction) correlators 

\begin{equation}
C_\Gamma(n_z) = \sum\limits_{n_x, n_y, n_t}
\braket{\mathcal{O}_\Gamma(n_x,n_y,n_z,n_t)
\mathcal{O}_\Gamma(\mathbf{0},0)^\dagger}
\label{eq:momentumprojection}
\end{equation}

\noindent
of all possible $J=0,1$
local isovector operators $\mathcal{O}_\Gamma(x) = \bar q(x) \Gamma \frac{\vec{\tau}}{2} q(x)$ have been calculated at temperatures above $T_c$
up to 380 MeV with the domain wall fermions \cite{R,C}. The operators as
well as their  $SU(2)_L \times SU(2)_R$ and $U(1)_A$ transformation
properties are presented in Table~\ref{tab:ops}.
\begin{table}
\center
\begin{tabular}{cccll}
\hline\hline
 Name        &
 Dirac structure &
 Abbreviation    &
 \multicolumn{2}{l}{
 } \\\hline
\textit{Pseudoscalar}        & $\gamma_5$                 & $PS$         & \multirow{2}{*}{$\left.\begin{aligned}\\ \end{aligned}\right] U(1)_A$} &\\
\textit{Scalar}              & $\mathds{1}$               & $S$          & &\\\hline
\textit{Axial-vector}        & $\gamma_k\gamma_5$         & $\mathbf{A}$ & \multirow{2}{*}{$\left.\begin{aligned}\\ \end{aligned}\right] SU(2)_A$}&\\
\textit{Vector}              & $\gamma_k$                 & $\mathbf{V}$ & & \\
\textit{Tensor-vector}       & $\gamma_k\gamma_3$         & $\mathbf{T}$ & \multirow{2}{*}{$\left.\begin{aligned}\\ \end{aligned}\right] U(1)_A$} &\\
\textit{Axial-tensor-vector} & $\gamma_k\gamma_3\gamma_5$ & $\mathbf{X}$ & &\\
\hline\hline
\end{tabular}
\caption{
Bilinear operators and their transformation properties. This classification assumes propagation in $z$-direction. The
 index $k$ denotes the components $1,2,4$, \textit{i.e.} $x,y,t$.}
\label{tab:ops}
\end{table}

\begin{figure}
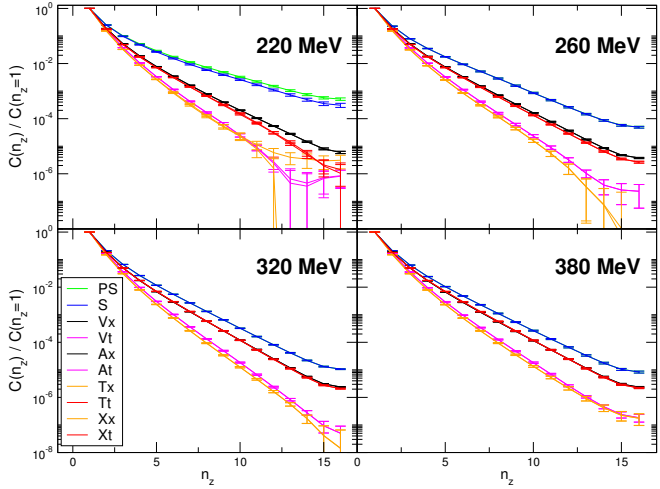

  \centering
  \includegraphics[scale=0.32]{{{1}}}
  \caption{
    Normalized spatial correlators.
  }
  \label{fig:corrs}
\end{figure}

Figure \ref{fig:corrs} shows the spatial correlators normalized to 1
at $n_z=1$ for all operators in Table~\ref{tab:ops}.
The argument  $n_z$ is proportional to the dimensionless product
$zT$.  
We observe three distinct multiplets:
\begin{eqnarray}
E_1: & \qquad PS \leftrightarrow S \label{eq:e1} \\
E_2: & \qquad V_x \leftrightarrow T_t \leftrightarrow X_t \leftrightarrow A_x \label{eq:e2} \\
E_3: & \qquad V_t \leftrightarrow T_x \leftrightarrow X_x \leftrightarrow A_t. \label{eq:e3}
\end{eqnarray}
$E_1$ is the Pseudoscalar-Scalar multiplet connected
by the $U(1)_A$ symmetry.
The $E_2$ and $E_3$ multiplets contain however
some operators that are  connected by neither
$SU(2)_L \times SU(2)_R$  nor $U(1)_A$
transformations. Below we demonstrate that the symmetries
responsible for emergence of the $E_2$ and $E_3$ multiplets are
$SU(2)_{CS}$ and $SU(4)$.

Consider the following dim=2 representations
of $SU(2)_{CS}$ that are defined in the Dirac bispinor space:
\begin{eqnarray}
R_1:\;
\{\gamma_1,-i\gamma_5\gamma_1,\gamma_5\}, \label{eq:R1} \\
R_2:\;
\{\gamma_2,-i\gamma_5\gamma_2,\gamma_5\}. \label{eq:R2}
\end{eqnarray}
Those differ from the representation
$\{\gamma_4,-i\gamma_5 \gamma_4,\gamma_5\}$ relevant
for $t$-direction correlators  by 
rotations.
The $R_1$ and $R_2$ $SU(2)_{CS}$ transformations of the quark fields
combine the following
operators from the $E_2$ multiplet into triplets:
\begin{eqnarray}
R_1: &\qquad
V_y \leftrightarrow T_t \leftrightarrow X_t, 
\\
R_2: &\qquad
V_x \leftrightarrow T_t \leftrightarrow X_t,
\end{eqnarray}
as well as  from the $E_3$ multiplet:
\begin{eqnarray}
R_1: &\qquad
V_t \leftrightarrow T_y \leftrightarrow X_y,
\\
R_2: &\qquad
V_t \leftrightarrow T_x \leftrightarrow X_x.
\end{eqnarray}
Given the $S_2$ symmetry ($x \leftrightarrow y$ permutations)  one obtains
the following $S_2 \times SU(2)_{CS}$  multiplets
\begin{align}
(V_x,V_y,T_t,X_t); \quad (V_t,T_x,T_y,X_x,X_y).
\end{align}
The degeneracy between $\mathbf{V}$ and $\mathbf{A}$ requires the 
$SU(2)_L \times SU(2)_R$ symmetry. Extending $SU(2)_{CS}$ to $SU(4)$
one arrives at the following multiplets of the isovector operators:
\begin{align}
(V_x,V_y,T_t,X_t,A_x,A_y); \; (V_t,T_x,T_y,X_x,X_y,A_t).
\label{eq:su4multiplet}
\end{align}
$S_2 \times SU(4)$ multiplets include in addition the isoscalar partners of
operators.

\begin{figure}
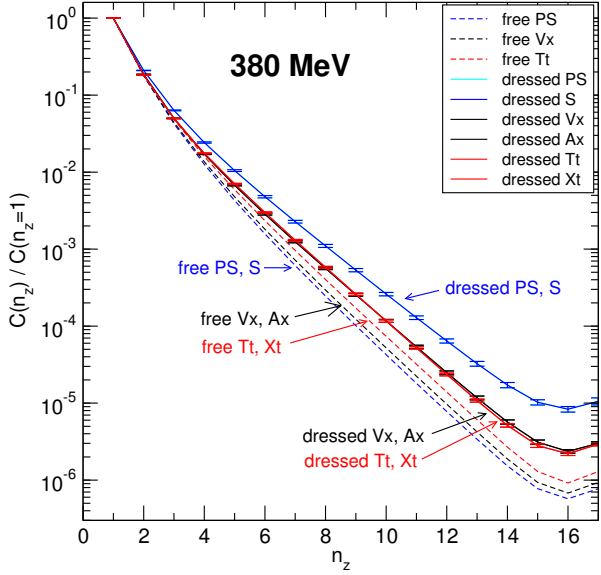

  \centering
  \includegraphics[scale=0.45]{{{2}}}
  \caption{
    $E_1$ and $E_2$ multiplets
    (\protect{\ref{eq:e1}}-\protect{\ref{eq:e2}}) for interacting
    (\textit{dressed}) and non-interacting (\textit{free}) calculations
    at $T$= 380~MeV.
  }
  \label{fig:e2}
\end{figure}

While the $U(1)_A$ and $SU(2)_L \times SU(2)_R$ symmetries are "exactly"
restored at temperatures above 220 MeV, the $SU(2)_{CS}$ and $SU(4)$ 
multiplets are only approximate. A degree of degeneracy at the highest
available temperature 380 MeV can be deduced from the Fig. \ref{fig:e2}
where we show correlators of the $E_1$ and $E_2$ multiplets in detail.
The remaining $SU(2)_{CS}$ and $SU(4)$ breaking
is at the level of 5\%. 
 We
also show there correlators calculated with the noninteracting quarks (abbreviated as "free"). 

The slopes of the PS and S correlators are substantially smaller than for the free quark-antiquark pair. This can 
happen only if the system represents a bound  meson-like state \cite{T}. In the free
quark case the slope is determined by twice of the lowest Matsubara frequency 
because of the antiperiodic boundary conditions for quarks in time direction.
If the quark-antiquark sytem is bound, and  of the bosonic nature the periodic boundary conditions (for bosons) do allow the slope to be smaller. For the 
$J=1$ correlators the difference of slopes of dressed and free correlators
is smaller than for the $J=0$ correlators, but is still visible.

\begin{figure}
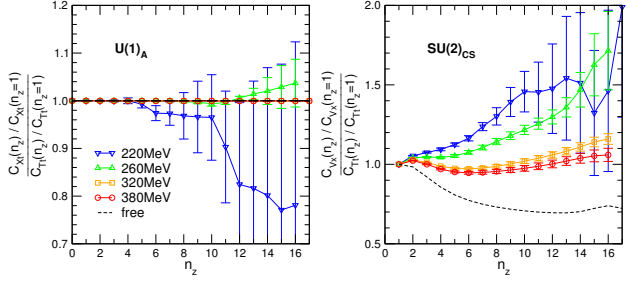

  \centering
  \includegraphics[scale=0.29]{{{4a}}}
  \includegraphics[scale=0.29]{{{4b}}}
  \caption{
    Ratios of normalized correlators ,
    that are related by $U(1)_A$ and $SU(2)_{CS}$ symmetries.
  }
  \label{fig:ratios}
\end{figure}

In order to see a tendency of the $U(1)_A$ restoration and of
$SU(2)_{CS}$ and $SU(4)$ emergence
we show  on Fig. \ref{fig:ratios} a ratio of the correlators from the multiplet
$E_2$ that are connected by the corresponding transformations   at different temperatures. We also show  a ratio calculated with the free noninteracting quarks.
We observe that upon increasing the temperature the correlators approach the
$SU(2)_{CS}$-symmetric limit and not the free quark limit as is
prescribed by the asymptotic freedom at high T (i.e. by the renormalization group flow equation calculated with perturbative $\beta$-function). This is a striking observation. We plan to verify this behaviour at higher temperatures.

\section{Implications}
We conclude that our lattice results are consistent with
emergence of the $SU(2)_{CS}$ and $SU(4)$ symmetries by
increasing temperature. The correlation functions do not seem
to approach the free quark limit.

These results have a direct implication on the nature of the
degrees of freedom in the high T phase. Emergence of the
$SU(2)_{CS}$ and $SU(4)$ symmetries rules out the possibility that 
elementary objects are deconfined asymptotically free quarks.
Instead such elementary objects should be chiral quarks
bound by the pure chromo-electric field that is not accompanied
by the magnetic effects (!), a kind of a string. Such a schematic
construction automatically incorporates the $SU(2)_{CS}$ and $SU(4)$ 
symmetries, as it follows from Sec. 4. Absence of the magnetic
effects implies that e.g. the spin-orbit force is absent and the
string with the left or with the right quark at the end has the same
energy \cite{GS}.

How should such a state of matter  be called? It is not
a plasma, because according to the standard definition
plasma is a system of free color charges where the chromo-electric 
field is Debye screened. According to our results it follows
that there are no free quarks and in addition it is the chromo-magnetic,
but not the chromo-electric field, is screend. So conditionally one could
call this matter  a stringy matter.

\bigskip
We acknowledge partial support from the Austrian Science Fund (FWF)
through the grant P26627-N27.

\end{document}